\def\PRL{{Phys. Rev. Lett.} }

\documentclass{ws-ijqi}

\catchline{}{}{}{}{}

\title{Numerical characterization of atmospheric effects on an Earth-Space quantum communication channel.}

\author{N. Antonietti \footnote{nicolo.antonietti@polito.it} , M. Mondin}

\address{Dipartimento di
         Elettronica, Politecnico di Torino, Corso Duca degli Abruzzi, 24,  (TO) - Italy
             }

\author{ G. Brida,
 M. Genovese \footnote{genovese@inrim.it}.}

\address{INRIM; strada delle Cacce 91, 10135 Torino, Italy}

\begin{document}

\maketitle
\begin{history}
\received{30-05-2006}
\end{history}

\begin{abstract}
Quantum communication in free space is the next challenge of
telecommunications. Since we want to determine the outcome of
quantum communication by means of single photons, we must
understand how a single photon interacts with the atmosphere. In
this brief article, some simulation results for realistic and
generic atmospheric conditions are reported, a related experiment
is considered and its results are described and discussed.
\end{abstract}

\keywords{entangled states;  quantum
communication.}

\section{Introduction}

Quantum Key Distribution (QKD) has now reached substantially a
commercial level having been realized up to many tens kilometer
both in fiber and open space \cite{gis}.

Nowadays the frontier of QKD is the realization of a Earth-Space or
a Space-Space quantum communication channel \cite{zei,rar,giap,nos}.

This realization would be of utmost relevance both for quantum
key distribution \cite{gis}, since it would allow intercontinental
quantum transmissions, and for studies concerning foundations of
quantum mechanics \cite{zei,mg}. Thus, preliminary feasibility
studies have been performed showing its practical realizability.

In little more details, in ref. \cite{rar} a BB84 scheme was
studied for earth-Space communication. By considering gaussian
optics, a 15 dB loss was attributed to diffraction, whilst
aerosol loss was considered of secondary relevance (0.04-0.06
dB) for  transmission with clear sky and from high elevation above
sea. Altogether atmospheric losses were estimated to be about 2-5
dB. On the other hand, the security level for quantum transmission
was estimated to be 40 dB, lowering to 10 dB if the eavesdropper
(conventionally dubbed Eve) had technologies for intercepting
selectively an eventual multi-photon component.

In another study, ref. \cite{zei}, a 6.5 dB loss was estimated by
considering optics and finite quantum efficiency of detectors. The
limit for secure quantum transmission was estimated to be at 60 dB
loss. Here atmospheric losses were estimated to be around 1 dB.

Finally, a more systematic study was performed in \cite{gil} by
using a program of US Airforce. Nevertheless, no detail was given
of the code and few results were really presented (atmospheric
losses were estimated to be 1 dB).

Effectively, daylight and open space transmission above 10 km
\cite{10} and at a distance of up to 23 km \cite{rz} were
performed, and an European collaboration is preparing an
experiment at Canary islands for a distance up to 120 km
\cite{tom}.

All these theoretical and experimental studies  guarantee the
feasibility of ground-space channel. Nevertheless, the analysis
of atmospheric effects is rather incomplete and is far from
considering various realistic atmospheric situations that could be
met during a real transmission. Even in the very general review
of ref. \cite{gil}, only few results are presented, and with small
detail.

Thus, a detailed analysis of atmospheric effects in various
realistic situations would be of the utmost relevance.

Purpose of this paper is to describe a work addressed to reach a
precise characterization of atmospheric effects on a quantum
communication channel by using a well tested simulation program.

To investigate this topic we have used a free source library for
radiative transfer calculations named libRadtran\cite{libradtran}.
This library can solve the radiative transfer equations, set some
input parameters and exploit the HITRAN\cite{hitran} database, which is a
high-resolution atmospheric parameters database.\\
In our simulations we can determine what part of the solar radiation
is the direct downward irradiance. However, we are still not able to say
anything about the possible photon depolarization (whose precise
estimation is still in progress).

Various parameters can influence the atmospheric effects on the
photon transmission, as, for instance, aerosols, pressure,
temperature, air density, precipitations, cloud composition,
humidity, chemical components. As a first step in order to
evaluate their relevance in various meteorological conditions,
here we present some preliminary results obtained by varying one
of them at time.

\section{Real experiments}
\label{sec:real_experiments}
\subsection{Different distributions of aerosols}
\label{subsec:aerosols}

In libRadtran, a database of aerosols distributions can be found. It
has been written according to ref. \cite{shettle:1989}. There, four
aerosols distributions for four different environment conditions are
described (rural, maritime, urban, tropospheric).\\

For our first analysis, the atmospheric conditions are selected to be
in summer, at midlatitudes, according to ref. \cite{afgl:1986} and
the source irradiance is chosen in accordance to ref.
\cite{kato:1999}; both databases are present in libRadtran. \\
For these conditions the downward direct irradiance at the Earth
surface with the source at the zenith outside the whole atmosphere
has been evaluated. We chose the visible wavelength because this is
the range used in quantum communications and is not affected by
strong absorption as UV and IR bands; actually the exact range is
from 256 nm to 1010.320 nm.

\begin{figure}[h]
\begin{center}
  \scalebox{0.8}{\includegraphics{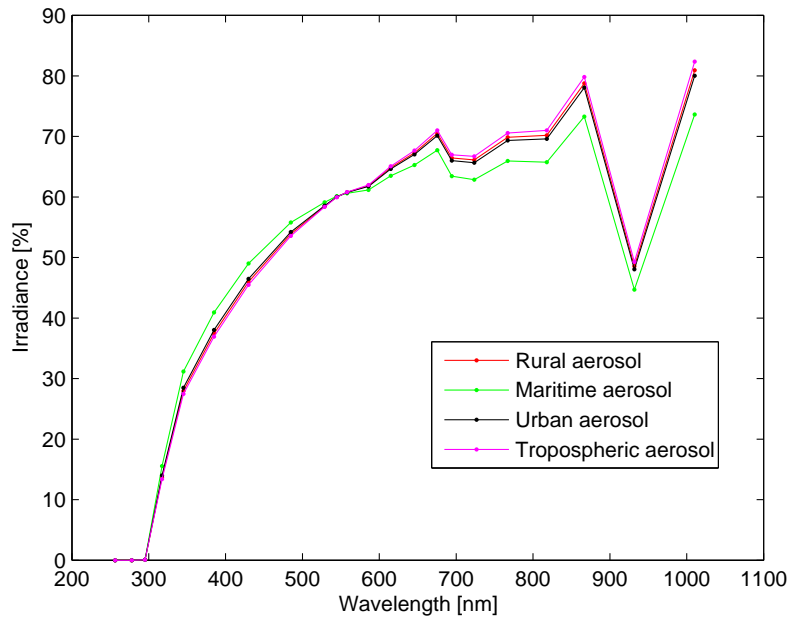}}\\
  \caption{Ratio of direct downward irradiance with respect to the source
  irradiance, in the visible range, for different aerosols
  conditions}\label{transmissivity_shettle_aerosols}
\end{center}
\end{figure}

A first result of this analysis is that this quantity is largely
independent on the aerosol type, as it can be observed in figure
(\ref{transmissivity_shettle_aerosols}).

Ut can also be observed that the best range for communications is
roughly from 700 nm to 900 nm. In this range, the fraction source
light which gets across the atmosphere without any interaction
with the atmosphere is about 75\% (i.e. a photon has a 75\%
probability to get across the model of atmosphere we have built
without interacting with it).

\subsection{Different temperature profile}
\label{subsec:temp_prof}

\begin{figure}[h]
\begin{center}
  \scalebox{0.8}{\includegraphics{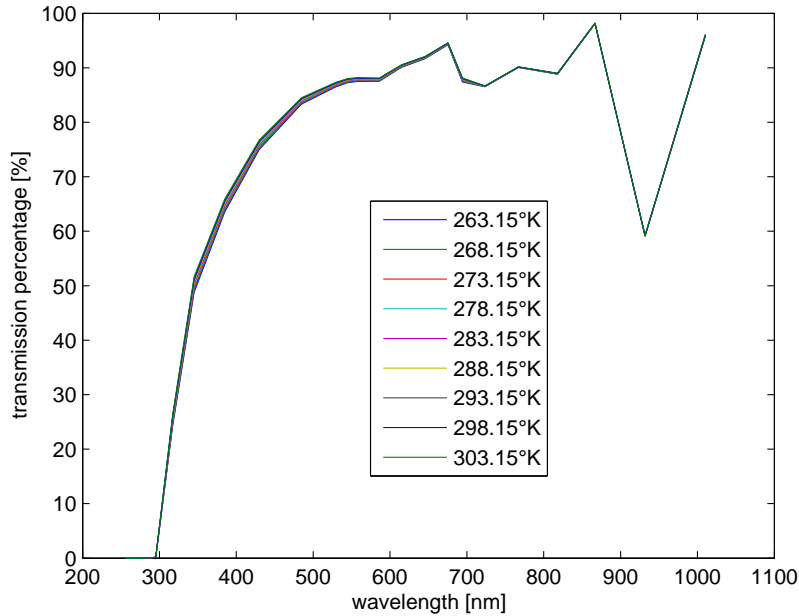}}\\
  \caption{Ratio of direct downward irradiance with respect to the source
  irradiance, in the visible range, for different temperature profiles}
  \label{transmissivity_temperature}
\end{center}
\end{figure}

In the considered atmosphere database, the temperature increases fairly
linearly from the ground level value $T_{0}$ up to 15 km, where it assumes
a given value $T_{15}$. $T_{0}$ is actually a free parameter, and we
have varied its value from -10$^o$C up to
30$^o$C, with steps of 5$^o$C. The air density is modified automatically, by the program, according to the perfect gas law.  Above 15 km, the parameters have been left unchanged, and no aerosols presence has been considered. The obtained results are depicted in figure
(\ref{transmissivity_temperature}).

Once again, it is possible to observe that the transmissivity is not affected by
the above modifications; and the best range for the communications
is roughly from 700 nm to 900 nm as well.

\subsection{Different humidity profiles}
\label{subsec:hum_prof}

\begin{figure}[h]
\begin{center}
  \scalebox{0.8}{\includegraphics{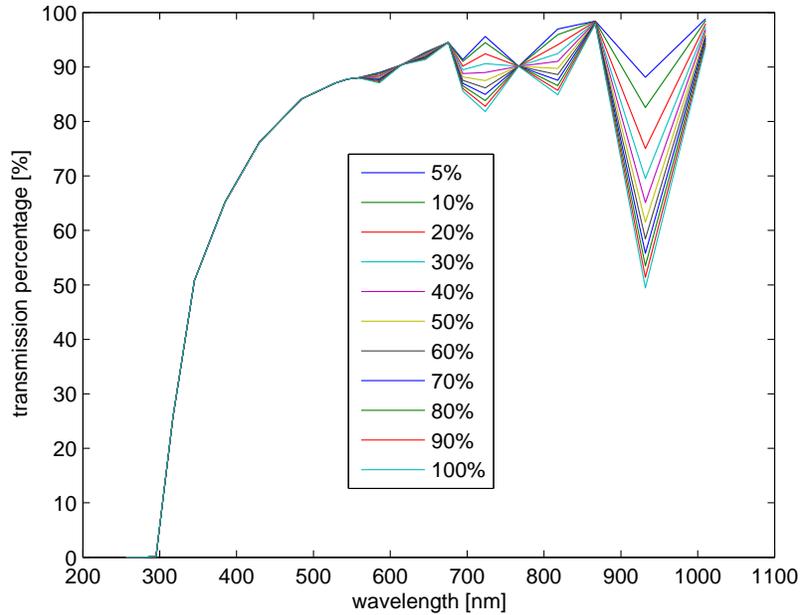}}\\
  \caption{Ratio of direct downward irradiance with respect to the source
  irradiance, in the visible range, for different relative humidity profiles}
  \label{transmissivity_humidity}
\end{center}
\end{figure}

In order to observe the effect of humidity, a further modification has been
added to the atmospheric
conditions of ref. \cite{afgl:1986}. This time, the relative
humidity has been set as a constant value in the first 15 km of the
atmosphere. The values are 5\% and from 10\% to 100\% with steps of
10\%. No aerosols have been considered in this configuration. In
figure (\ref{transmissivity_humidity}), the obtained results can be observed.

Also in this scenario, the most advantageous range for
communication is from 700 nm to 900 nm, but in this case we can
observe some differences among different humidity conditions. As
it can be observed from figure (\ref{transmissivity_humidity}),
the HITRAN\cite{hitran} database, being based on measurements
performed with wave-packets and not with photons, has, in some
regions, a somewhat low resolution with respect to what would be
needed for our application. As a consequence, some low-resolutions
effects, like those observed in the 700 nm to 1000 nm region, can
arise. The absorption in the range between 800 nm and 1000 nm is
water vapour dependent and is strongly affected by its presence.
These preliminary results point out the necessity of a more
precise analysis with an increased resolution in wave length.

\subsection{Presence of clouds}
\label{subsec:cloud}

\begin{figure}[h]
\begin{center}
  \scalebox{0.8}{\includegraphics{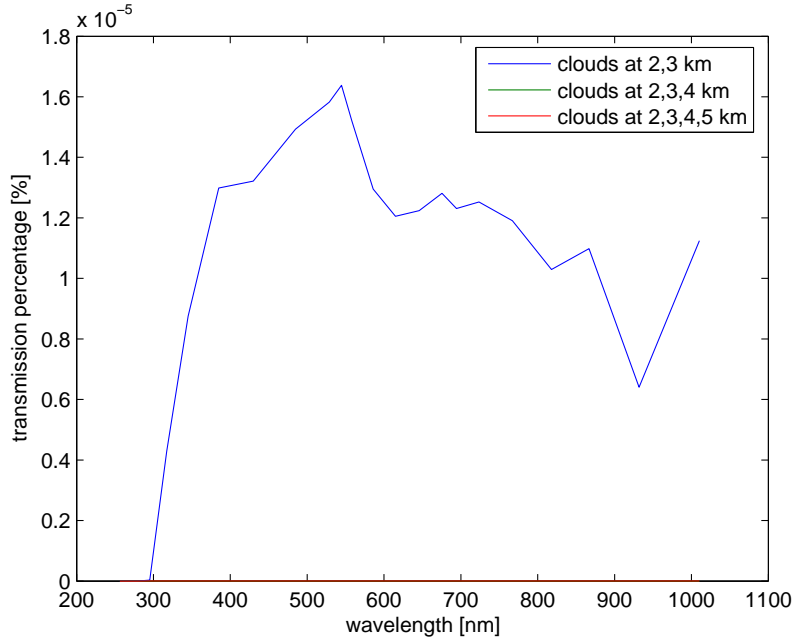}}\\
  \caption{Ratio of direct downward irradiance with respect to the source
  irradiance, in the visible range, for different cloud configurations}
  \label{transmissivity_clouds}
\end{center}
\end{figure}

In order to study the possibility of establish a quantum
communication channel, the presence of clouds has to be considered
as well. In order to do this, we have added clouds to the
atmosphere \cite{afgl:1986} without aerosols. The model we use
describes the clouds as two-dimensional objects, without depth. We
considered different cloud configurations, inserting clouds at 2,3
km, at 2,3,4 km and at 2,3,4,5 km. In our model, every cloud has a
liquid water content of 1$gm^{-3}$\ and an effective droplet
radius of 10$\mu m$. The results are presented in figure
(\ref{transmissivity_clouds}).

It can be immediately observed that the presence of clouds
seriously harms the quantum communication. Only the configuration
with two layers of clouds at 2 km and 3 km gives a transmission
percentage value different from zero, actually the transmissivity
values are in the order -5, and therefore not suitable for our
application. Notice that the selected cloud configurations belong
to the set contained within the libRadtran database, but also
other configurations could be considered, if needed.

\subsection{Comparison between two extremely different conditions}
\label{subsec:diff_cond}

\begin{figure}[h]
\begin{center}
  \scalebox{0.8}{\includegraphics{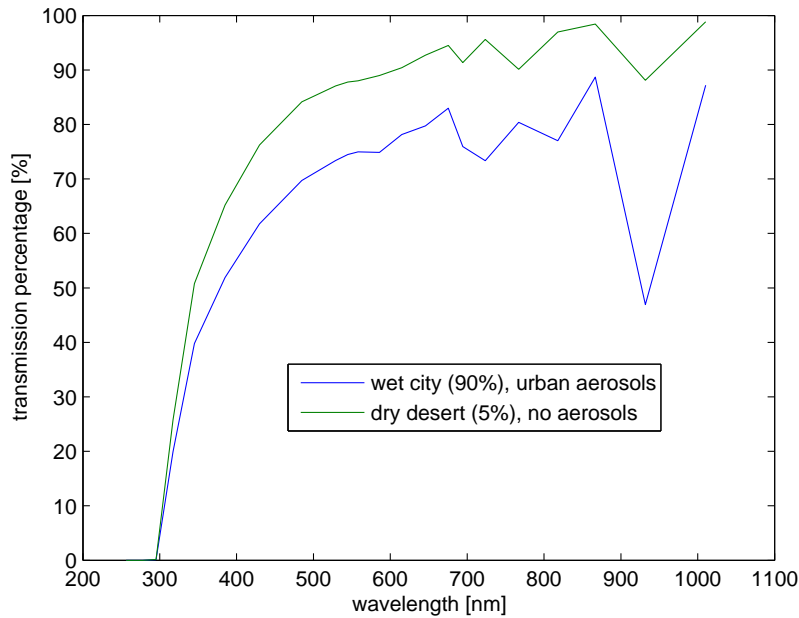}}\\
  \caption{Ratio of direct downward irradiance with respect to the source
  irradiance, in the visible range, for two extremely different environments}
  \label{transmissivity_citydesert}
\end{center}
\end{figure}

Finally we want to get an idea of how is transmissivity for two
extremely different conditions. On one side there is a city
environment with relevant aerosols concentrations and 90\% relative
humidity, on the other side a dry desert without aerosols. The results for these
two cases are depicted in figure (\ref{transmissivity_citydesert}).

As we could expect, a dry desert is a much better environment for
quantum communication than a humid city. Nevertheless, also in
this second case losses are not such to compromise the possibility
of realizing a secure QKD.

\section{Conclusions}
\label{sec:conclusions}

If we want to investigate atmospheric effects on a quantum channel,
with the purpose of predicting the possible results of an
experiment, we have to solve stochastic equations. In this paper we
have presented some preliminary results obtained by using the free source
library libRadtran. Our results show that a further deeper
analysis of atmospheric effects based on this approach could
effectively be a useful tool for predicting the performances of a
quantum communication channel in various realistic operative
meteorological situations.

\end{document}